\documentclass[%
 preprint,
 nofootinbib,
 amsmath,amssymb,
 aps,
 floatfix,
]{revtex4-1}

\usepackage[margin=1in]{geometry}
\usepackage{graphicx}
\usepackage{setspace}
\usepackage{color}
\usepackage{dcolumn}
\usepackage{bm}
\usepackage{subcaption}

\begin{document}


\begin{center}
$J/\psi$-$p$ Scattering Length from the Total and Differential Photoproduction Cross Sections
\end{center}
\author{Lubomir~Pentchev}
\email[Corresponding author: ]{pentchev@jlab.org}
\affiliation{Thomas Jefferson National Accelerator Facility, Newport News, Virginia 23606, USA}
\author{Igor~I.~Strakovsky}
\affiliation{The George Washington University, Washington, D.C. 20052, USA}

\date{\today}

\begin{abstract}
The $J/\psi$-$p$ scattering length, $\alpha_{J/\psi p}$, can be extracted from the $J/\psi$ photoproduction cross section near threshold using the Vector Meson Dominance (VMD) model to relate the reaction $\gamma p \to J/\psi p$ to $J/\psi p \to J/\psi p$. Such estimates based on experimental data result in values for $|\alpha_{J/\psi p}|$, which are much lower than most of the theoretical predictions. In this work, we study the relations between the different results, depending on the use of the total or the differential cross sections, and the method of extrapolating the data to threshold in the case of a low-statistics data sample, such as the near threshold $J/\psi$ photoproduction dataset. We estimate a range for $|\alpha_{J/\psi p}|$ of $0.003$ to $0.025$~fm as extracted from experimental data within the VMD model and discuss possible reasons for such lower values compared to the theoretical results.
\end{abstract}

\maketitle


\clearpage
\mbox{~}
\section{Introduction}

There is a special interest to study the $J/\psi$ - nucleon interaction because of the small size of charmonium that can be used to probe the internal structure of the nucleon. 
Experimentally, the charmonium-nucleon interaction can be investigated using $J/\psi$ photoproduction within the Vector Meson Dominance (VMD) model. The {\it near threshold} exclusive reaction, $\gamma p \to J/\psi p$, can be used to extract the $J/\psi$-$p$ scattering length as the final state particles are produced with a small momentum in the center-of-mass (CM) frame. The applicability of the VMD model in this case requires special attention. Actually, near threshold, we are not dealing with a $J/\psi$ in equilibrium, but rather with a ``young" $c\bar{c}$ system~\cite{Feinberg:1980yu}. In such an interaction, more time is needed for the slow heavy quarks to reach equilibrium, {\it i.e.,} to form the final on-mass-shell vector meson.  

In Ref.~\cite{Strakovsky_jpsi}, the $J/\psi$-$p$ scattering length is estimated using the recent measurement of the total $J/\psi$ photoproduction cross section near threshold from the GlueX Collaboration~\cite{prl_gluex}. Within the VMD model, the total $\gamma p \to J/\psi p$ cross section is related to the total $J/\psi p \to J/\psi p$ cross section and, at threshold, to the scattering length $\alpha_{J/\psi p}$ by~\cite{Titov}: 
\begin{equation}
    \sigma^{\gamma p}(s_{thr}) = 
    \frac{\alpha\pi}{\gamma_\psi ^2}\frac{q_{\psi p}}{k_{\gamma p}}
    \cdot \sigma^{\psi p}(s_{thr}) =
    \frac{\alpha\pi}{\gamma_\psi ^2}\frac{q_{\psi p}}{k_{\gamma p}}
    \cdot 4\pi \alpha_{J/\psi p}^2~.
    \label{eq:method2}
\end{equation}
Here $k_{\gamma p}$ and $q_{\psi p}$  are the momenta in the CM of the initial and final state particles, respectively, and $\gamma_\psi$ is the photon - $J/\psi$ coupling constant obtained from the $J/\psi\to e^+e^-$ decay width.  The above equation is taken at the threshold energy, where $s = s_{thr} = (M + m)^2$ with $M$ and $m$ being the masses of the $J/\psi $ and proton, respectively. When approaching threshold, $q_{\psi p}$ approaches $0$. Therefore, with this method, the derivative of the cross section as a function of $q_{\psi p}$, for $q_{\psi p}\to 0$, is estimated from the data and then related to the scattering length by Eq.~(\ref{eq:method2}). In Ref.~\cite{Strakovsky_jpsi}, this is done by fitting the data with an odd-power polynomial function. The result for the absolute value of the scattering length is $|\alpha_{J/\psi p}| = (0.00308\pm 0.00055~(stat.) \pm 0.00045~(syst.)$)~fm. This value is much smaller than the values obtained from the processing of experimental data by other methods and theoretical estimates (see References in~\cite{Strakovsky_jpsi}).

\section{Scattering Length from Differential Cross Sections}

In order to use measurements of the differential photoproduction cross section, ${d\sigma^{\gamma p}}/{dt}$, to estimate the scattering length, first, we will establish the relation between the total and differential cross sections at threshold. The total cross section is determined as an integral over $t$ in the interval $t_{min}(s)\geq t \geq t_{max}(s)$: 
\begin{equation}
    \sigma^{\gamma p}(s) = \int _{t_{min}(s)}^{t_{max}(s)} \frac{d\sigma^{\gamma p}}{dt}(s,t) dt\;,
\end{equation}
with $t_{min,max}(s) = M^2 - 2k_{\gamma p}(E_\psi \pm q_{\psi p})$, where $E_\psi^2 = q_{\psi p}^2 + M^2$. When approaching threshold $t_{min}\to t_{max}$, and we have for the above integral:
\begin{equation}
    \sigma^{\gamma p}(s_{thr}) = \Delta t\frac{d\sigma^{\gamma p}}{dt} (s_{thr}, t_{thr}) = 4 q_{\psi p} k_{\gamma p}\frac{d\sigma^{\gamma p}}{dt}
    (s_{thr},t_{thr})\;,
    \label{eq:totdiff}
\end{equation}
where $\Delta t = |t_{max} - t_{min}| = 4 q_{\psi p}k_{\gamma p}$ and $t_{thr} = t_{min}(s_{thr}) = t_{max}(s_{thr}) = -M^2 m/(M + m)$. The above equation
relates the total and differential cross sections at threshold. Combining Eq.~(\ref{eq:totdiff}) and Eq.~(\ref{eq:method2}), we obtain:
\begin{equation}
    \frac{d\sigma^{\gamma p}}{dt}(s_{thr}, t_{thr}) = 
    \frac{\alpha \pi}{\gamma_\psi ^2}\frac{\pi }{k_{\gamma p}^2}\cdot \alpha_{J/\psi p}^2~.
    \label{eq:method2d}
\end{equation} 

The extrapolation of the cross sections to the point of $t\to t_{thr}$ or $s\to s_{thr}$ ($q_{\psi }\to 0$) is a key problem in determination of the scattering length. From an experimental point of view, the total cross section is more suitable for extrapolation to the threshold than the differential cross section, since the latter case needs higher statistics.
Thus, the GlueX Collaboration reported the differential cross section, ${d\sigma^{\gamma p}}/{dt}$, as function of $t$ in an energy range of $10 \;-\; 11.8$~GeV with an average energy of $10.7$~GeV~\cite{prl_gluex}, which corresponds to $q_{\psi p} = 0.95$~GeV/$c$, while the lowest-energy data point for the total cross section is at a four-times smaller $q_{\psi p}$ of $0.23$~GeV/$c$, {\it i.e.,} much closer to the threshold. Therefore, to utilize the differential cross section data, one needs to know the energy dependence of $d\sigma^{\gamma p}/{dt}(s)$ in order to extrapolate it reliably to threshold.

In Ref.~\cite{Brodsky}, the asymptotic behavior of $J/\psi$ photoproduction near threshold is studied using dimensional scaling. Due to the OZI rule, the $J/\psi -p$ interaction is mediated predominantly by gluons. In this approach, the differential cross section of $J/\psi$ photoproduction is analyzed in the framework of a two-component model
with 2- and 3-gluon exchange. Each component describes the contribution to the differential cross section in terms of the quark-gluon parton model:
\begin{equation}
    \frac{d\sigma^{\gamma p}_i}{dt} = {\cal N}_i\,
    (1-x)^{2n_{s}}\cdot F_i^2(t)~,
    \label{eq:brodsky2}
\end{equation}
where $n_s$ is the number of spectators in the proton target not participating in the process; $n_s = 1$ and 0 for the two- and three-gluon exchange channels ($i = 2, 3$), respectively. $F_i(t)$ is a proton form factor that takes into account the fact that the outgoing quarks recombine into the final proton after the gluon emission, for which we use the dipole form: 
$F_i(t) = (1 - t/1.3~(\rm GeV^2))^{-2}$~\cite{prl_gluex}. 
The scale variable $x$ near threshold is chosen as $x = (2mM + M^2)/(s - m^2)$. The total cross section is determined as an integral over $t$ in the interval $t_{min}(s)\geq t \geq t_{max}(s)$. 
We find the normalization constants ${\cal N}_i$ from a fit of the GlueX~\cite{prl_gluex} and SLAC~\cite{SLAC} total cross-section data. 
\begin{figure}[ht]
\vspace{-0.5cm}
\centering
{
    \includegraphics[width=0.8\textwidth,keepaspectratio]{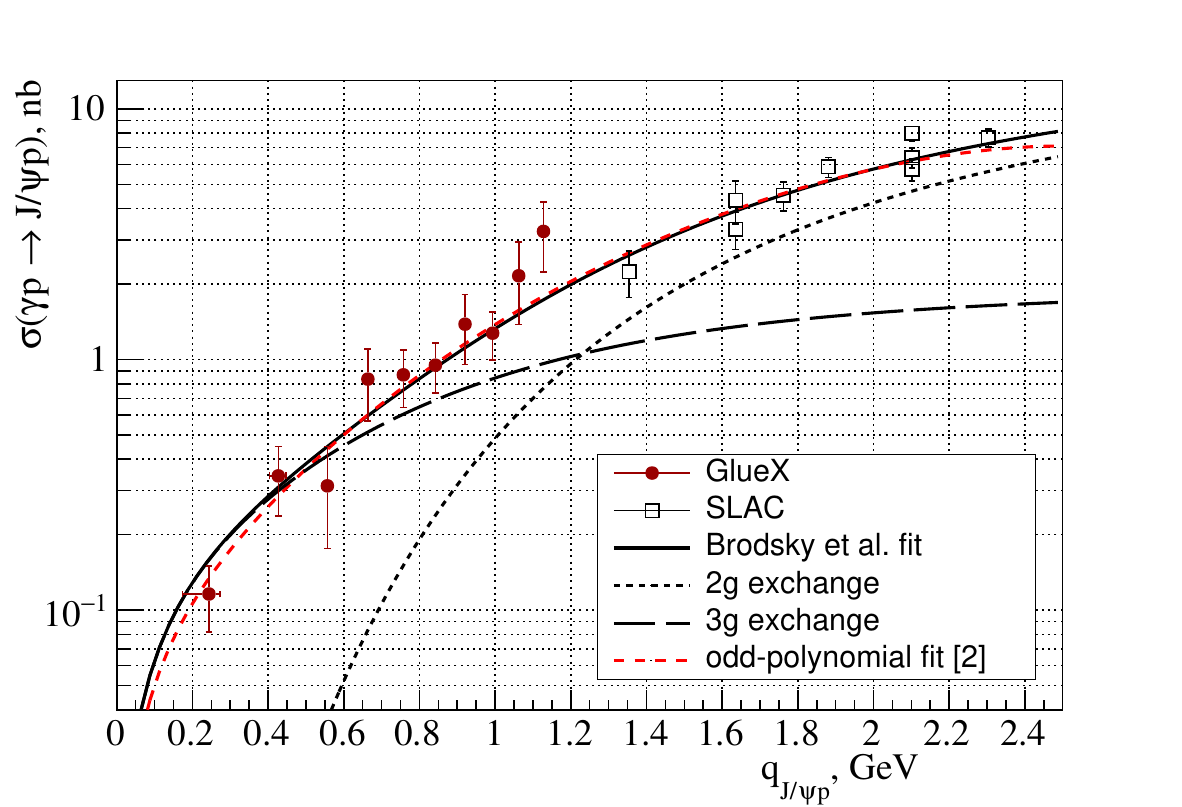}
}

\centerline{\parbox{0.70\textwidth}{
  \caption[] {\protect\small 
  The exclusive $J/\psi$ photoproduction cross sections from GlueX~\protect\cite{prl_gluex} and SLAC~\protect\cite{SLAC} fitted  with an incoherent sum of two- and three-gluon exchange contributions of Eq.~(\protect\ref{eq:brodsky2}) and an odd-power polynomial as in Ref.~\protect\cite{Strakovsky_jpsi}, as a function of the momentum of the outgoing $J/\psi$ in the CM. The SLAC total cross sections are obtained from the differential cross sections using the procedure in Ref.~\protect\cite{prl_gluex}.}
  \label{fig:xsec_fit} } } 
\end{figure}

The fit results are shown in Fig.~\ref{fig:xsec_fit} with the separate contributions from the two- and three-gluon exchange as a function of the CM momentum in the final state as the dotted and long-dashed curves, respectively. Their incoherent sum is depicted by the solid curve. For completeness, the fitting by an odd-power polynomial~\cite{Strakovsky_jpsi} is shown by the short-dashed curve. 
Similarly to Ref.~\cite{prl_gluex}, we find that the three-gluon exchange dominates in the GlueX energy region. Therefore, within this model, the differential cross section near threshold does not depend on the energy as follows from Eq.~(\ref{eq:brodsky2}) for $n_s = 0$, but only on $t$. Besides other important consequences, this means that we can use measurements of the $t$-dependence at energies away from threshold to predict the cross section near threshold.

\begin{figure}[ht]
\centering
{
    \includegraphics[width=0.6\textwidth,keepaspectratio]{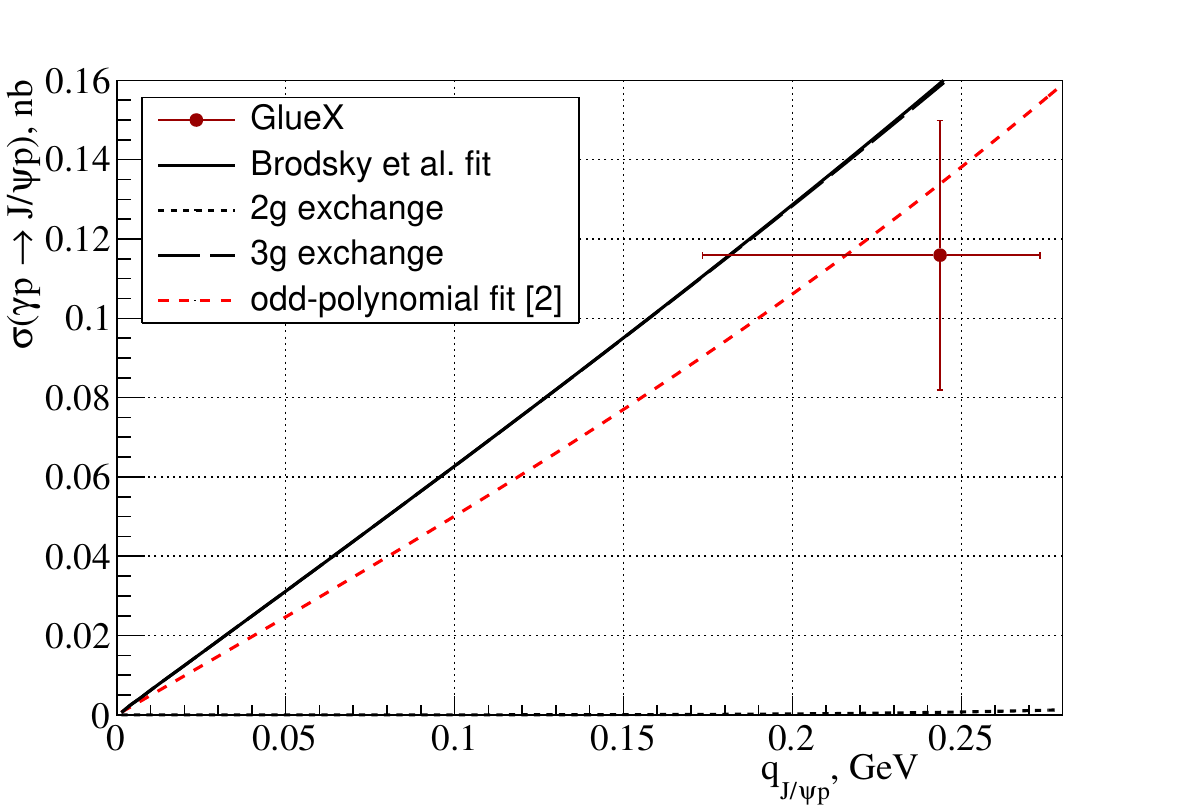}
}

  \centerline{\parbox{0.70\textwidth}{
    \caption[] {\protect\small
    Fig.~\protect\ref{fig:xsec_fit} zoomed near threshold. 
    The three-gluon exchange curve (long dash) coincides with the total (solid) Brodsky \textit{et al.} fit~\protect\cite{Brodsky}.}
  \label{fig:xsec_zoom} } } 
\end{figure}

Fitting the GlueX differential cross section data~\cite{prl_gluex} with an exponential function,
\begin{equation}
    \frac{d\sigma^{\gamma p}}{dt}(t) = A \cdot e^{b(t - t_{min})}~,
    \label{eq:expfit}
\end{equation}
results in a slope of $b = (1.67\pm 0.35)$~GeV$^{-2}$ and $A = (1.83\pm 0.32)$~nb/GeV$^2$. Using $t_{min} = -0.44$~GeV$^2$ for the energy of $10.7$~GeV and $t_{thr} = -2.23$~GeV$^2$, we estimate for the right-hand side of Eq.~(\ref{eq:totdiff}) a $q$-slope of
\begin{equation}
    4q_{\psi p}k_{\gamma p}\frac{d\sigma^{\gamma p}}{dt}(t = t_{thr}) = q_{\psi p}\cdot(0.71\pm0.35)~{\rm nb/GeV}~.
    \label{eq:rhs}
\end{equation}
For the left-hand side of Eq.~(\ref{eq:totdiff}), it is found in Ref.~\cite{Strakovsky_jpsi} using an odd-power polynomial fit to the GlueX total cross section, a slope of $(0.46\pm 0.16)$~nb/GeV. If we calculate the derivative at $q_{\psi p} = 0$ of the three-gluon exchange function that was used to fit the total cross section data (see Fig.~\ref{fig:xsec_zoom}), we get a value of $(0.64\pm 0.09)$~nb/GeV. Both values are in agreement with Eq.~(\ref{eq:rhs}).
Thus we have verified based on Eq.~(\ref{eq:totdiff}) and assuming energy independence of the differential cross section that when using the total or the differential cross sections from the GlueX measurements, the results for the scattering length are consistent.

In Ref.~\cite{Vanderhaeghen_jpsi}, a method to extract the real and imaginary parts of the $J/\psi$-$p$ forward amplitude $T^{\psi p}(t = 0)$ is presented. It is based on fits of the ``elastic'' $\gamma p\to J/\psi p$ and inelastic $\gamma p\to c\bar{c} X$ total cross sections and the forward differential cross section $d\sigma^{\gamma p}/dt(t = 0)$.  The imaginary part is extracted using the optical theorem and parameterizations of the total cross sections. Then the real part is obtained using dispersion relations with one subtraction. The subtraction constant is defined by the real part of the amplitude at threshold. Using VMD, the latter is constrained by the forward differential cross section at threshold:
\begin{equation}
    \frac{d\sigma^{\gamma p}}{dt}(s_{thr}, t = 0) = 
    \frac{\alpha\pi}{\gamma_\psi ^2}\frac{q_{\psi p}^2}{k_{\gamma p}^2}\cdot
    \frac{d\sigma^{\psi p}}{dt}(s_{thr}, t = 0) =
    \frac{\alpha\pi}{\gamma_\psi ^2}\frac{1}{64\pi s_{thr} k_{\gamma p}^2}\cdot |T^{\psi p}(s_{thr}, t = 0)|^2~,
\end{equation}
as the imaginary part vanishes there. On the other hand, the forward angle scattering amplitude at threshold is related to the $J/\psi$-$p$ 
scattering length, $\alpha_{J/\psi p}$, as: 
\begin{equation} 
    T^{\psi p}(s_{thr}, t = 0) = 8\pi \sqrt{s_{thr}}\cdot \alpha_{J/\psi p}~.
\end{equation} 
Therefore, the scattering length is obtained practically from:
\begin{equation}
    \frac{d\sigma^{\gamma p}}{dt}(s_{thr}, t = 0) = 
    \frac{\alpha\pi}{\gamma_\psi^2}\frac{\pi}{k_{\gamma p}^2}\cdot \alpha_{J/\psi p}^2~.
    \label{eq:method1}
\end{equation}
For the purpose of our studies, we will associate Eq.~(\ref{eq:method1}) with the results in Ref.~\cite{Vanderhaeghen_jpsi}, despite the more complicated fitting procedure described above. The left-hand side is not a quantity that can be measured directly, as it requires both extrapolation in energy to threshold and extrapolation in $t$ from the physical region $(t_{min}, t_{max})$ to the nonphysical point $t = 0$. The differential cross section data closest to threshold that is used in Ref.~\cite{Vanderhaeghen_jpsi} is from the SLAC~\cite{SLAC} measurements above 13~GeV, which corresponds to $q_{\psi p} > 1.35$~GeV/$c$, which is far away from the energy threshold of $q_{\psi p} =0$. The best fit results in a value for the scattering length of $|\alpha_{J/\psi p}| = (0.046 \pm 0.005)$~fm \cite{Vanderhaeghen_jpsi}.

We notice the difference between Eq.~(\ref{eq:method1}) related to Ref.~\cite{Vanderhaeghen_jpsi}, and Eq.~(\ref{eq:method2d}) that stems from the method of using the total cross section~\cite{Strakovsky_jpsi}. In the former case, the differential cross section $d\sigma^{\gamma p}/dt$ is taken at $t = 0$, while in the latter case it is at $t = t_{thr} = -2.23~{\rm GeV}^2$. This is a result of the fact that when the total cross section approaches the threshold energy ($q_{\psi p} = 0$), $t$ approaches $t_{thr}$. On the other hand, when using the differential cross section, we can extrapolate not only to the energy at threshold, but also in $t$ to $t = 0$. 
\begin{table}[hp]
\begin{ruledtabular}
\begin{tabular}{lll}
\textrm{Extrapolated data (method)}&
\textrm{$|\alpha_{J/\psi p}|$ $10^{-3}$~fm} &
\textrm{~~Reference}\\
\colrule
$\sigma^{\gamma p}(s_{thr})$,  GlueX \protect\cite{prl_gluex} (odd-polynomial fit) & $3.08\pm 0.55$ & \protect\cite{Strakovsky_jpsi} \\
$\sigma^{\gamma p}(s_{thr})$,  GlueX \protect\cite{prl_gluex} (3g-exchange model)  & $3.64\pm 0.26$ & this work \\
$d\sigma^{\gamma p}/dt(s_{thr},t_{thr})$, GlueX \protect\cite{prl_gluex} $10.7$~GeV (energy independence) & $3.83\pm 0.98$ & this work \\
$d\sigma^{\gamma p}/dt(s_{thr},0)$, SLAC \protect\cite{SLAC} $>13$~GeV (global fit) & $ 46\pm 5 $& \protect\cite{Vanderhaeghen_jpsi} \\
$d\sigma^{\gamma p}/dt(s_{thr},0)$, GlueX \protect\cite{prl_gluex} $10.7$~GeV (energy independence) & $24.5\pm 3.9$ & this work \\
\colrule
\textrm{Theoretical models (year)}& & \\
\colrule
Photoproduction via open-charm channel (2020) & $0.2-3$ & \protect\cite{OpenCharm} \\
QCD multipole expansion (2020) & $200-2000$ & \protect\cite{Krein} \\
Lattice QCD (2019)& $200-700$ & \protect\cite{Sugiura}\\
Lattice QCD (2019)& small & \protect\cite{Skerbis}\\
Lattice QCD (2006)& $710 \pm 480$ & \protect\cite{Yokokawa:2006td}\\
Multipole expansion, LE QCD theorem (2005) & $370$ & \protect\cite{Sibirtsev:2005ex} \\
QCD sum rules (1999) & $100$ & \protect\cite{Hayashigaki:1998ey} \\
Gluonic van der Waals interaction (1997) & $250$ & \protect\cite{Brodsky:1997gh} \\
$q\bar{q}$ Green's function, non-relativistic gluonic interaction (1997) & $12$ & \protect\cite{Shevchenko:1996ch} \\
Heavy-quarkonia gluonic interaction, LE QCD theorem (1992) & $50$ & \protect\cite{Kaidalov:1992hd} \\
\end{tabular}
\end{ruledtabular}
\caption{Results for the absolute value of the $J/\psi$-$p$        scattering length obtained from $J/\psi$ photoproduction using different datasets and extrapolating methods as described in the text (only statistical uncertainties are shown) - top. 
These results are compared to the theoretical calculations~\cite{OpenCharm,Krein,Sugiura,Skerbis,Yokokawa:2006td,Sibirtsev:2005ex,Hayashigaki:1998ey,Brodsky:1997gh,Shevchenko:1996ch,Kaidalov:1992hd} - bottom.
The lattice results of Ref.\cite{Skerbis} ``are roughly in agreement with the predictions for almost noninteracting nucleon and $J/\psi$''.
}
\label{tab:comp}
\end{table}

We can study quantitatively the difference between the two results 
using the fit (Eq.~(\ref{eq:expfit})) of the GlueX differential cross section assuming energy independence near threshold, this time for $t \to 0$, as well. We have $d\sigma^{\gamma p}/dt(t = t_{thr}) = (0.093\pm 0.045)$~nb/GeV$^2$, while $d\sigma^{\gamma p}/dt(t = 0) = (3.8\pm 1.2)$~nb/GeV$^2$. In addition, the SLAC cross section measurements used in Ref.~\cite{Vanderhaeghen_jpsi} are done away from threshold ($q_{\psi p} > 1.35$~GeV/$c$) and the best fit extrapolates them to $d\sigma^{\gamma p}/dt(t = 0) \approx 10$~nb/GeV$^2$ at threshold (see Fig.~3 in Ref.~\cite{Vanderhaeghen_jpsi}). Thus, the differential cross section method of Ref.~\cite{Vanderhaeghen_jpsi} effectively uses a value for $d\sigma^{\gamma p}/dt$ that is two orders of magnitude bigger than the value that corresponds (from Eq.~(\ref{eq:totdiff})) to the total cross section method used in Ref.~\cite{Strakovsky_jpsi}. This explains the order of magnitude difference between the two results for the scattering length that enters quadratically.

In Table~\ref{tab:comp} (top), we summarize the results for the $J/\psi$-$p$ scattering length obtained from data as discussed in this work. The top three lines represent the method of using the total cross section for two fitting functions and also utilizing the differential cross section at $t = t_{thr}$. The next two lines represent the use of the differential cross section extrapolated to $t = 0$ from the original work~\cite{Vanderhaeghen_jpsi} and also from the recent GlueX data assuming energy independence near threshold. For comparison, in Table~\ref{tab:comp} (bottom), we give results of the theoretical calculations as they were cited in Ref.~\cite{Strakovsky_jpsi} including also some recent works. Generally, the scattering length extracted from data is much lower than most of the theoretical predictions.

\section{Discussions and Outlook}

The GlueX Collaboration has recently studied $J/\psi $ photoproduction off the proton near threshold ~\cite{prl_gluex}. The proximity of the {\it total} cross section data to threshold allowed for an estimate the absolute value of the $J/\psi$-$p$ scattering length within the VMD model~\cite{Strakovsky_jpsi}.  This result agrees well with our determination of the scattering length using the GlueX {\it differential} cross section data away from the threshold and assuming its energy independence: $d\sigma/dt(t_{thr},s) = d\sigma/dt(t_{thr},s_{thr})$. Using the same assumption but for $t = 0$, we have explained the difference between the results of Refs.~\cite{Strakovsky_jpsi} and \cite{Vanderhaeghen_jpsi}. Such energy independence near threshold has important implications in extrapolating the measured cross sections to threshold and requires further experimental studies. 
In this work, we relied on the model of Ref.~\cite{Brodsky} for which in the case of three-gluon exchange, the differential cross section $d\sigma/dt$ depends only on $t$ and not on $s$. Despite such a model assumption, generally, we expect the dependence of $d\sigma/dt$ on energy to be much weaker than on $t$, which makes the presented extrapolation method more reliable when working with low statistics. 

We found that within the VMD model, when using the GlueX photoproduction data, the $J/\psi$-$p$ scattering length is estimated to be in the range of $(3 - 25)\times 10^{-3}$~fm. 
One of the interpretations of such small values compared to most of the theoretical predictions (see Table~\ref{tab:comp}) came from  Feinberg's consideration back in the 1970s~\cite{Feinberg:1980yu}. 
The measured scattering length is very small because of the small size of the ``young'' $J/\psi$. 
As we deal with the $J/\psi$ created by the photon near threshold, it is not formed completely. 
The transverse $c$-$\bar{c}$ separation of the ``young'' $J/\psi$ estimated to be $r_{J/\psi} \approx 1/m_c = 0.13$~fm 
(see Fig.1 in \cite{Brodsky}, $m_c$ is the charm quark mass),  is smaller than that for the ``old'' on-shell $J/\psi$, $R_{J/\psi} \approx 0.3$~fm. 
Consequently, the $J/\psi$-nucleon interaction is suppressed by a factor  $\sim r^2_{J/\psi}/R^2_{J/\psi}$, compared to the VMD prediction. 
In other words, the $J/\psi$ has not had sufficient time to be formed completely and we observe a weaker interaction in the near-threshold photoproduction.

The ``young'' $J/\psi$ corresponds to the off-shell one, respectively the
ratio $R_{J/\psi}/r_{J/\psi}$ reflects its virtuality.
Thus, the $J/\psi $ off-shellness between the $\gamma\to J/\psi$ vertex and the $J/\psi p$ scattering has a significant impact
and may affect both the  $\gamma\to J/\psi$ coupling and the $J/\psi p$ scattering amplitude.
For more quantitative estimations of the effect of the VMD assumption, as discussed in Ref.~\cite{Strakovsky_jpsi}, we refer to the evaluation of the cross section of the $J/\psi $ photoproduction in the peripheral model \cite{Boreskov77}.
There, a strong energy dependence of the suppression factor close to threshold was observed, reaching a value of about 5 at threshold.
In another approach, it is argued in Ref.~\cite{Kopeliovich} that, as the color factor for the charmonium is 1/9 compared to 8/9 for the open charm production, 
fluctuations of the photon into open charm are preferable than into a $J/\psi $.
In Ref.~\cite{OpenCharm} the open-charm channel $\Lambda_c \bar{D}^{(*)}$ is used 
to calculate the near-threshold cross section of the $J/\psi $ photoproduction, free of the VMD assumption. 
Their result for the $J/\psi$-$p$ scattering length of $0.2-3$~$10^{-3}$~fm is at the lower end of the theoretical predictions and close to the results of this work and Ref.~\cite{Strakovsky_jpsi} when using the total cross-section data (see Table~\ref{tab:comp}).

The difference between Eq.~(\ref{eq:method1}) and Eq.~(\ref{eq:method2d}), discussed above, where the threshold differential cross-sections are taken at $t = t_{thr}$ or $t = 0$, originates from the use of the photoproduction to extract the $J/\psi$-$p$ scattering length and thus, such uncertainty is also related to the applicability of the VMD model. Indeed, for the $J/\psi p$ elastic scattering the two equations are equivalent as $t_{thr} = 0$. There is a significant additional uncertainty when extrapolating the differential cross-section data to the non-physical point at $t = 0$. Therefore, in an attempt to take into account such uncertainties, the result of this work should be considered as an estimate of the {\it range} of values for the scattering length as extracted from the experimental data within the VMD model.

\textbf{Acknowledgments}

We thank Michael Ryskin and Alexander Titov for useful remarks and continuous interest in the paper and Daniel Carman for valuable comments. This work was supported in part by the U.S. Department of Energy, Office of Science, Office of Nuclear Physics, under Award No.~DE–-SC0016583 and Contract No.~DE-–AC05-–06OR23177.

\bibliography{LPentchev_jpsi_sl.bib}

\end{document}